\documentclass[cameraready]{Interspeech}

\title{Predictive Directional Selective Fixed-Filter Active Noise Control for Moving Sources via a Convolutional Recurrent Neural Network}

\author[affiliation={1}]{Boxiang}{Wang}
\author[affiliation={1}]{Zhengding}{Luo$^*$}
\author[affiliation={2}]{Dongyuan}{Shi}
\author[affiliation={1}]{Junwei}{Ji}
\author[affiliation={2}]{Xiruo}{Su}
\author[affiliation={1}]{Woon-Seng}{Gan}

\address{
    $^1$ School of Electrical and Electronic Engineering, Nanyang Technological University, Singapore \\
    $^2$ Center of Intelligent Acoustics and Immersive Communications, Northwestern Polytechnical University, China 
    \thanks{$^*$Corresponding author: Zhengding Luo. The code will be available at \href{https://github.com/Wang-Boxiang/PD-SFANC}{https://github.com/Wang-Boxiang/PD-SFANC}.}
}

\email{boxiang001@e.ntu.edu.sg, luoz0021@e.ntu.edu.sg}


\usepackage{comment}
\usepackage{amssymb}
\usepackage{amsmath,graphicx,hyperref}
\usepackage{booktabs}
\usepackage{url}
\usepackage{tcolorbox}
\usepackage{tabularx}
\usepackage{multirow}
\usepackage{lineno}

\begin{document}

\maketitle

\begin{abstract}
\vspace*{-0.2cm}
Directional Selective Fixed-Filter Active Noise Control (D-SFANC) can effectively attenuate noise from different directions by selecting the suitable pre-trained control filter based on the Direction-of-Arrival (DoA) of the current noise. However, this method is weak at tracking the direction variations of non-stationary noise, such as that from a moving source. Therefore, this work proposes a Predictive Directional SFANC (PD-SFANC) method that uses a Convolutional Recurrent Neural Network (CRNN) to capture the hidden temporal dynamics of the moving noise and predict the control filter to cancel future noise. Accordingly, the proposed method can significantly improve its noise-tracking ability and dynamic noise-reduction performance. Furthermore, numerical simulations confirm the superiority of the proposed method for handling moving sources across various movement scenarios, compared to several representative ANC baselines.


\end{abstract}

\vspace*{-0.2cm}
\section{Introduction}
\vspace*{-0.2cm}
Active noise control (ANC) is an advanced technique that can effectively attenuate low-frequency noise through the principle of sound destructive interference, in which a secondary source generates anti-noise with equal amplitude and opposite phase to the unwanted noise~\cite{elliott1993active,kuo1999active}. Compared with passive methods that rely on bulky barriers, ANC provides a more compact and effective solution. Consequently, it has been widely applied in various applications, particularly for enhancing speech intelligibility and preserving acoustic comfort in noise-polluted environments~\cite{zhang2020deep,cheer2019application,samarasinghe2016recent,xiao2023spatially}. However, most existing ANC systems have been developed for stationary sources. In practice, the positions of noise sources are often time-varying, such as those generated by vehicles, drones, and vacuum cleaners, necessitating ANC systems tailored for moving sources~\cite{kuo2010active}.

\begin{figure}[!t]
\centering
\includegraphics[width=3.1in]{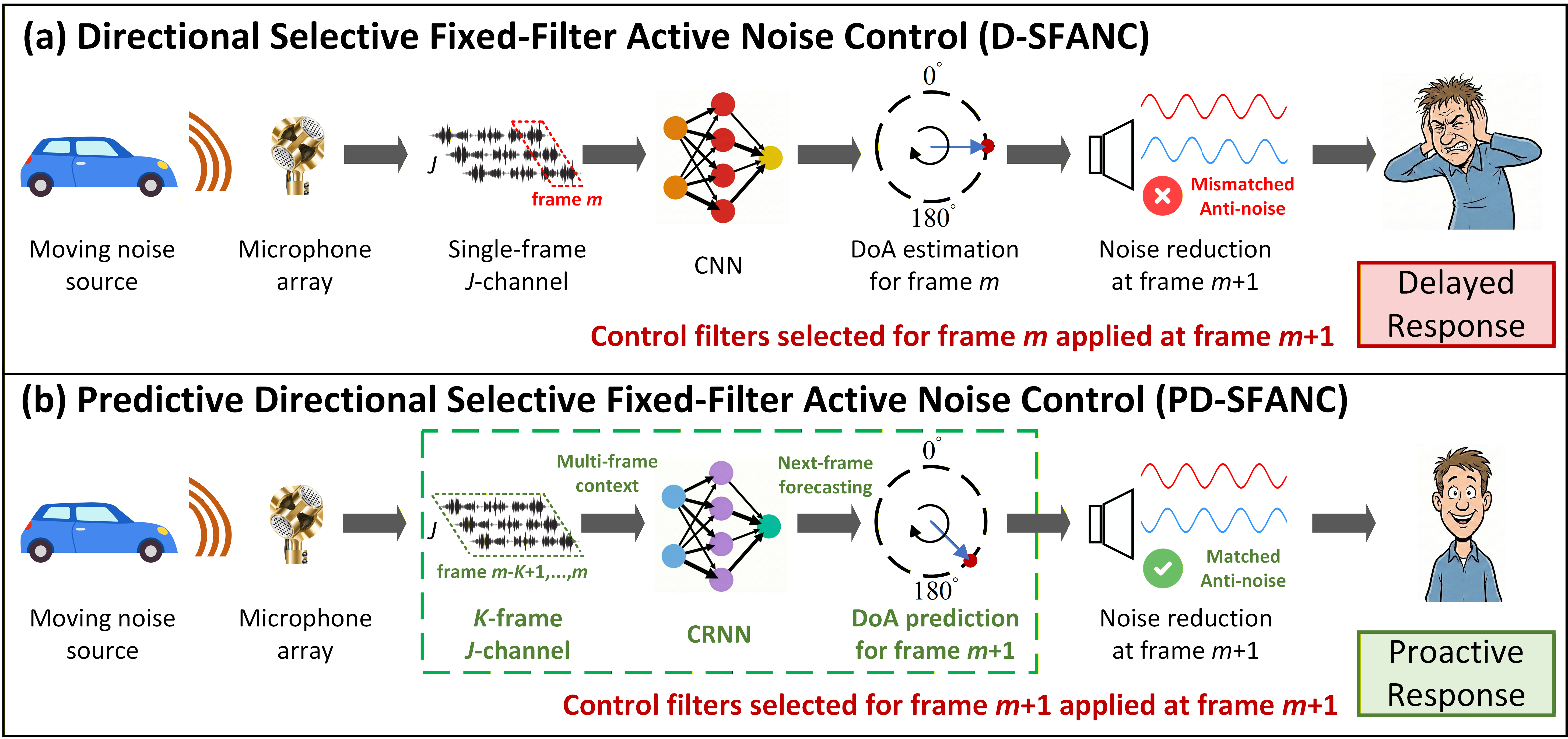}\vspace*{-0.3cm}
\caption{Comparison between (a) directional SFANC and (b) the proposed predictive directional SFANC.}
\label{fig_1}
\vspace*{-0.6cm}
\end{figure}

To attenuate moving noise sources, traditional ANC systems often utilize adaptive algorithms, such as the filtered-reference least mean squares (FxLMS) algorithm, to update the control filter in real time~\cite{omoto2002behavior,guldenschuh2014least,ho2021time}. However, these algorithms often suffer from slow convergence and are at risk of divergence~\cite{yang2023active,li2023augmented,liu2024study}. Furthermore, the requirement to place an error microphone at the target location imposes physical constraints~\cite{wang2025deep,ji2025self}. To alleviate these problems, the selective fixed-filter ANC (SFANC) method has been proposed to select the most suitable pre-trained control filter for various noise types~\cite{shi2020feedforward,luo2024real,wang2025transferable}. Nevertheless, these approaches overlook the spatial characteristics of the noise source, which significantly affect ANC performance~\cite{liebich2018direction,zhang2023time,xie2024cognitive,toyooka2025active}. To address this, the Directional SFANC (D-SFANC) method has been proposed, incorporating Direction-of-Arrival (DoA) information into the filter selection process~\cite{su2024spatial,luo2025doa,wangdirectional}. However, as shown in Fig.~\ref{fig_1}, D-SFANC fails to respond promptly and continues to use the control filter selected for the previous frame. This lag leads to degraded noise reduction during source transitions. Although the recent dynamic factor graph-based SFANC (DFG-SFANC) demonstrated that control filter pre-selection can improve performance~\cite{su2025co}, it relies on traditional signal processing techniques where several key parameters require empirical tuning.

To address these challenges, this paper proposes a Predictive Directional SFANC (PD-SFANC) method that employs a Convolutional Recurrent Neural Network (CRNN) to predict the DoA of moving noise sources in real time. By leveraging temporal context across multiple consecutive frames, the CRNN effectively captures the evolution of the source trajectory. This predictive capability enables the system to proactively select the most suitable control filter for the upcoming frame, ensuring stable and superior noise reduction with minimal latency. Furthermore, all CRNN parameters are learned automatically, eliminating the need for manual parameter tuning and significantly simplifying the system design. Notably, the proposed framework is designed for single-source scenarios, which is consistent with the system formulation in this work.

The remainder of this paper is organized as follows. Section~\ref{section2} details the proposed CRNN-based PD-SFANC method. Section~\ref{section3} evaluates the performance of the proposed algorithm through numerical simulations. Section~\ref{section4} concludes the paper.

\begin{figure}[!t]
\centering
\includegraphics[width=3.3in]{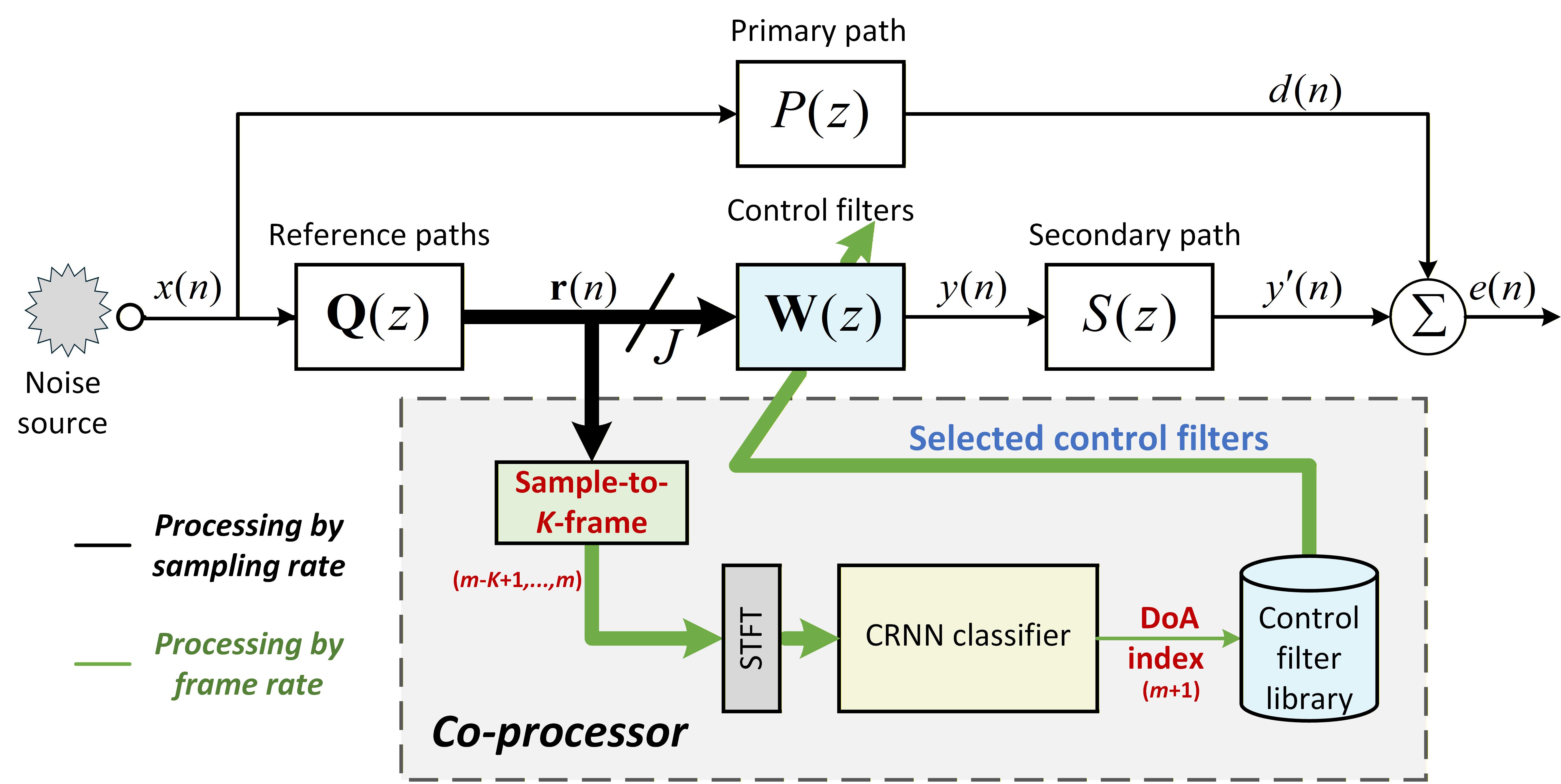}\vspace*{-0.3cm}
\caption{Block diagram of the predictive directional SFANC.}
\label{fig_2}
\vspace*{-0.5cm}
\end{figure}

\vspace*{-0.3cm}
\section{Predictive Directional SFANC}
\vspace*{-0.2cm}
\label{section2}
We introduce the PD-SFANC method to address the core challenge of delayed response for moving source noise control. As shown in Fig.~\ref{fig_2}, a CRNN running on the co-processor performs DoA prediction and selects the most suitable control filter for the \textit{upcoming} frame. In parallel, real-time noise control is executed at the sampling rate to achieve proactive noise control.

\vspace*{-0.2cm}
\subsection{Pre-trained control filter library}
\vspace*{-0.2cm}
Prior to the online execution of PD-SFANC, a control filter library is pre-trained to accommodate noise sources at various DoAs. Assume a discrete grid of DoAs for the noise source, denoted as ${\theta _v} \in \{ {\theta _1},\ldots,{\theta _V}\}$, where $V$ is the number of candidate angles. To alleviate the design complexity, at each DoA ${\theta _v}$, a control filter vector ${{\bf{w}}^{[{\theta _v}]}}$ is pre-trained using the FxLMS algorithm with broadband bandlimited white noise as
\begin{equation}
\setlength{\abovedisplayskip}{2pt}
\setlength{\belowdisplayskip}{2pt}
{{\mathbf{w}}^{[{\theta _v}]}}(n + 1) = {{\mathbf{w}}^{[{\theta _v}]}}(n) + \mu [{{\mathbf{r}}^{[{\theta _v}]}}(n)]'{e^{[{\theta _v}]}}(n),
\end{equation}
where $\mu$ is the stepsize, ${e^{[{\theta _v}]}}(n)$ is the error signal, and $[{{\mathbf{r}}^{[{\theta _v}]}}(n)]'$ is the filtered reference signal vector generated by passing the reference signal vector ${{\mathbf{r}}^{[{\theta _v}]}}(n)$ through the estimated secondary path $\hat s(n)$ as 
\begin{equation}
\setlength{\abovedisplayskip}{2pt}
\setlength{\belowdisplayskip}{2pt}
[{{\mathbf{r}}^{[{\theta _v}]}}(n)]' = \hat s(n)*{{\mathbf{r}}^{[{\theta _v}]}}(n).
\end{equation}

The resulting control filter vectors are stored in a library $[{{\mathbf{w}}^{[{\theta _v}]}}]_{v = 1}^V$ for deployment during online noise control.

\vspace*{-0.2cm}
\subsection{CRNN for moving source DoA prediction}
\begin{figure*}[!t]
\centering
\includegraphics[width=5.5in]{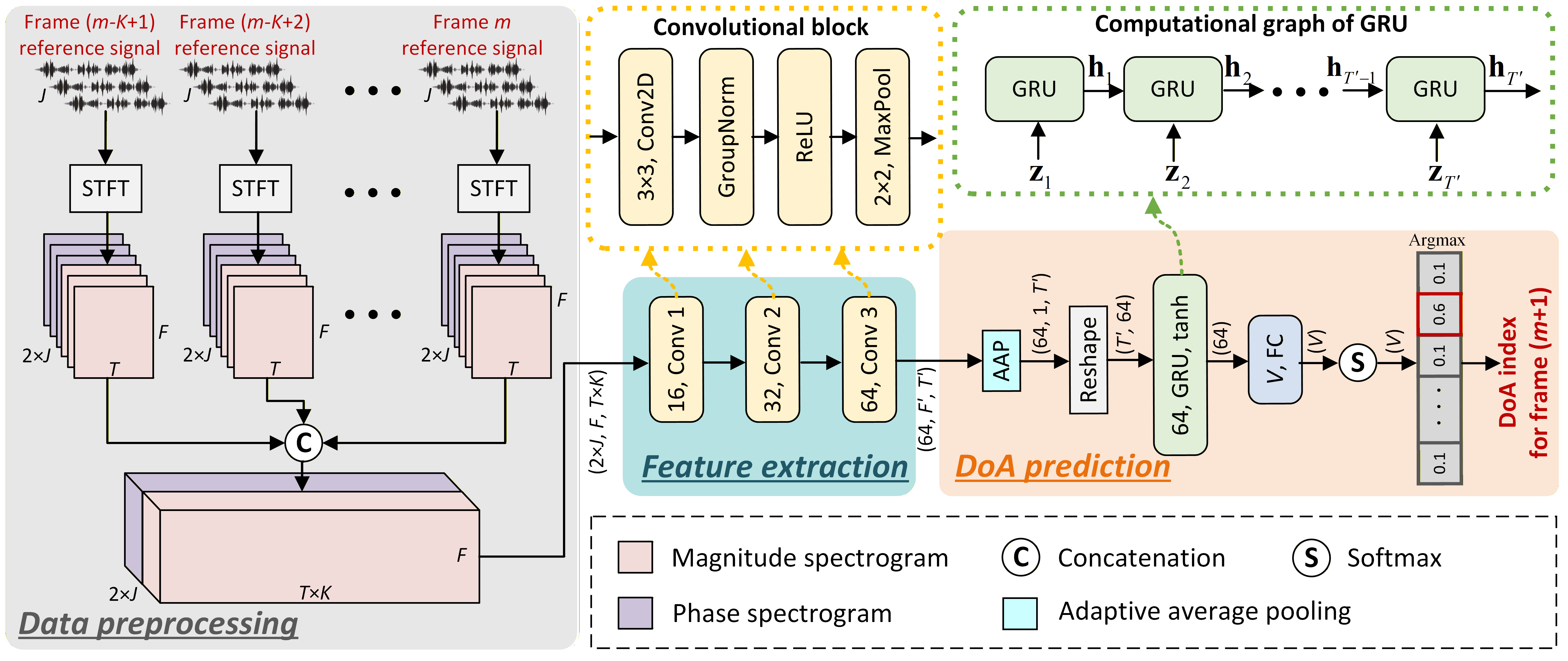}\vspace*{-0.3cm}
\caption{Proposed CRNN architecture for next-frame DoA prediction using multi-frame context.}
\label{fig_3}
\vspace*{-0.4cm}
\end{figure*}

\vspace*{-0.2cm}
CRNNs have been widely applied to sound source localization and moving source tracking tasks due to their ability to extract spatial features and capture temporal dependencies \cite{adavanne2019localization,li2023doa}. In this work, a CRNN is employed to forecast the next-frame DoA of a moving source through classification. The architecture of the proposed CRNN is shown in Fig.~\ref{fig_3}.

\vspace*{-0.2cm}
\subsubsection{Data Preprocessing}
\vspace*{-0.2cm}
To effectively exploit temporal dynamics, the CRNN input stacks $K$ consecutive frames of the $J$-channel reference signals. A short-time Fourier transform (STFT) is applied to the multichannel reference signal for each frame. In the STFT domain, the reference signal at the $j$-th microphone is defined as
\begin{equation}
\setlength{\abovedisplayskip}{2pt}
\setlength{\belowdisplayskip}{2pt}
R_j(m,t,f)=A_j(m,t,f)e^{i\phi_j(m,t,f)}, \quad j=1,\ldots,J,
\end{equation}
where $m$ denotes the frame index, $A_j(m,t,f)$ and $\phi_j(m,t,f)$ denote the magnitude and phase components, respectively. Here, $t=1,\ldots,T$ and $f=1,\ldots,F$ index the time frames and frequency bins, respectively, and $i = \sqrt{-1}$. The magnitude and phase spectrograms are concatenated along the channel dimension. Finally, the $K$ context frames are concatenated along the time axis, yielding an input tensor $\mathbf{R}\in\mathbb{R}^{2J\times F\times TK}$.

\vspace*{-0.2cm}
\subsubsection{CRNN Architecture}
\vspace*{-0.2cm}
To extract spatial features, the preprocessed input tensor is passed through three convolutional blocks. Each block consists of a two-dimensional (2D) convolutional layer, group normalization, rectified linear unit (ReLU) activation, and max pooling, with convolutions applied across the time-frequency dimensions. Adaptive average pooling is subsequently applied along the frequency axis to reduce dimensionality. The resulting feature map is reshaped as
\begin{equation}
\setlength{\abovedisplayskip}{2pt}
\setlength{\belowdisplayskip}{2pt}
\mathbf{z} = {\operatorname{Avg}}\left [\operatorname{CNN}(\mathbf{R})\right ]\in\mathbb{R}^{T'\times 64},
\end{equation}
where $\mathbf{z}=[\mathbf{z}_1\ldots\mathbf{z}_t\ldots\mathbf{z}_{T'}]^{\mathrm T}$, and $T'$ is the time-axis length of the pooled sequence. The feature map is then fed into a gated recurrent unit (GRU) layer, which fuses the extracted spatial features with inter-frame temporal dynamics. As shown in Fig.~\ref{fig_3}, the computation process of GRU is defined as
\begin{equation}
\setlength{\abovedisplayskip}{2pt}
\setlength{\belowdisplayskip}{2pt}
{\mathbf{h}}_t = {\operatorname{GRU}}({{\mathbf{z}}_t},{\mathbf{h}}_{t - 1})\in \mathbb{R}^{64},\quad t=1,\dots,T',
\end{equation}
where $\mathbf{h}_t$ denotes the hidden state at time $t$. By integrating information across consecutive frames, the GRU effectively models the temporal evolution of the DoA. Finally, a fully connected (FC) layer with softmax activation maps the final hidden state to the class probabilities of the $V$ DoAs, expressed as
\begin{equation}
\setlength{\abovedisplayskip}{2pt}
\setlength{\belowdisplayskip}{2pt}
\mathbf{\hat{p}} = \mathrm{Softmax} \left [ \mathrm{FC} ({\mathbf{h}}_{T'}) \right ]\in \mathbb{R}^{V},
\end{equation}
where $\mathbf{\hat{p}} = [{\hat p}_1\ldots{\hat p}_v\ldots{\hat p}_V]$. The predicted DoA index for the moving source in the next frame is given by 
\begin{equation}
\setlength{\abovedisplayskip}{2pt}
\setlength{\belowdisplayskip}{2pt}
\hat{v} = \mathop {\arg \max }\limits_{{v} \in \{ 1,\ldots,V\} } {\hat {p}_v}.
\end{equation}
\vspace*{-0.5cm}
\subsubsection{Loss function}
\vspace*{-0.2cm}
A cross-entropy loss function is employed to optimize the predicted DoA distributions, expressed as
\begin{equation}
\setlength{\abovedisplayskip}{2pt}
\setlength{\belowdisplayskip}{2pt}
\mathcal{L} = -\sum_{v=1}^{V} y_v \log(\hat{p}_v).
\end{equation}
where $y_v$ is the one-hot ground truth label. The CRNN is trained using the Adam optimizer \cite{kingma2014adam} to obtain the optimal parameters. 

\begin{table}[!t]
\caption{Pseudo-code of the predictive directional SFANC.}
\vspace*{-0.3cm}
\label{table1}
\scriptsize
\begin{tabularx}{\linewidth}{@{}X@{}}
\toprule
\textbf{Initialization:} The control filter vector is initialized to zero.\\
\textbf{Input:} $K$ consecutive frames of the $J$-channel reference signal. \\
\textbf{Note:} $\mathbf{r}_{m-k}$ denotes the $J$-channel reference signal in frame $m-k$.\\
\midrule
\textbf{While PD-SFANC is on:} \\ 
\textbf{\# Noise control in the real-time controller (sampling rate):}\\
\quad $y(n) = \mathbf{w}^{\mathrm T}(n)\mathbf{r}(n)$ \hfill$\triangleright$ Control signal.\\ 
\quad $e(n) = d(n) - s(n) \ast y(n)$ \hfill$\triangleright$ Real-time noise control.\\ 
\textbf{\# Control filter pre-selection in the co-processor (frame rate):}\\
\quad \textbf{for} $k=K-1,\ldots,0$ \textbf{do}\\
\qquad $\mathbf{R}_{m-k} \leftarrow \operatorname{Concat}\left[\left| \operatorname{STFT}(\mathbf{r}_{m-k}) \right|,\angle \operatorname{STFT}(\mathbf{r}_{m-k})\right]$ \\ 
\quad \textbf{end for}\\
\quad $\mathbf{R} \leftarrow \operatorname{Concat}\left(\mathbf{R}_{m-K+1},\ldots,\mathbf{R}_{m}\right)$ \\
\quad $\hat{v} = \operatorname{CRNN}(\mathbf{R})$ \hfill$\triangleright$ Next-frame DoA index prediction.\\ 
\quad $\mathbf{w}^\prime \leftarrow \mathbf{w}^{[\theta_{\hat{v}}]}$ \hfill$\triangleright$ Control filter pre-selection.\\
\textbf{\# Control filter update in the real-time controller (frame rate):}\\
\quad \textbf{if} $\mathbf{w} \neq \mathbf{w}^\prime$ \textbf{then}\\
\qquad $\mathbf{w} \leftarrow \mathbf{w}^\prime$ \hfill$\triangleright$ Control filter update for subsequent noise control.\\
\quad \textbf{end if}\\
\bottomrule
\end{tabularx}
\vspace*{-0.5cm}
\end{table}

\vspace*{-0.2cm}
\subsection{Proactive noise control}
\vspace*{-0.2cm}
During online operation, PD-SFANC achieves proactive noise control through a dual-module architecture consisting of a co-processor and a real-time controller. The co-processor (e.g., a mobile phone) executes the CRNN to pre-select the most appropriate control filter at the frame rate, while the real-time controller operates at the sampling rate to perform immediate noise cancellation. This coordinated design ensures delayless noise control by decoupling real-time processing from the latency introduced by the CRNN. The pseudo-code for the PD-SFANC procedure is presented in Table~\ref{table1}. 

Importantly, the system forecasts the noise source's movement: the selected control filter ${\mathbf{w}}^{[{\theta_{\hat v}}]}$ corresponds to the predicted DoA of the \textit{upcoming} frame, effectively eliminating latency during control filter transitions. Following a short $K$-frame cold-start period for context accumulation, the control filter is updated every frame without buffering delays. Furthermore, unlike traditional adaptive ANC algorithms, PD-SFANC eliminates reliance on feedback error signals for online filter adaptation, thereby improving response time and minimizing the risk of divergence. As a result, the system provides a practical solution for suppressing moving noise sources.

\begin{table}[!t]
\centering
\caption{Simulation parameters.}
\vspace*{-0.3cm}
\label{table2}
\scriptsize
\begin{tabular}{|c|c|c|}
\hline
\textbf{Variable} & \textbf{Definition} & \textbf{Value} \\
\hline
-- & Sampling rate & $16000$ Hz \\
$J$ & Number of reference microphones & $4$ \\
-- & Number of secondary sources & $1$ \\
-- & Number of error microphones & $1$ \\
-- & Control filter length & $1024$ \\
-- & Secondary path length & $256$ \\
$F$ & STFT frequency bins & $513$ \\
$T$ & STFT time frames & $64$ \\
$V$ & Number of DoA categories & $36$ \\
$K$ & Number of consecutive frames & $4$ \\
-- & Frame length & $0.5$ s \\
-- & Network input length& $2$ s \\
\hline
\end{tabular}
\vspace*{-0.4cm}
\end{table}

\vspace*{-0.4cm}
\section{Numerical Simulations}
\vspace*{-0.2cm}
\label{section3}
\subsection{Simulation setup}
\vspace*{-0.2cm}
The simulation parameters are summarized in Table \ref{table2}. Specifically, the reference microphone array is a tetrahedral arrangement of four cardioid microphones ($0.025$ $\mathrm{m}$ diameter) following the \textit{Sennheiser AMBEO VR Mic} geometry, which compactly captures the spatial information~\cite{kushwaha2023sound}. Owing to the small array aperture, the noise source is assumed to be in the far field.

The DoA of the noise source is defined by the azimuth angle with respect to the reference microphone array. A discrete DoA grid is constructed by uniformly sampling $[0, 360)^\circ$, with a resolution of $10^\circ$. At each DoA, a control filter is pre-trained with broadband noise up to $2$ kHz, covering the low-frequency band typically targeted by ANC systems \cite{wang2025transferable}. In total, $36$ control filters are pre-trained and stored in the library.

\vspace*{-0.4cm}
\subsection{Dataset construction}
\vspace*{-0.2cm}
The CRNN datasets are generated by convolving noise signals with simulated multichannel room impulse responses (RIRs). The noise signals include (i) synthesized bandlimited white noise with a random bandwidth under $2$ kHz and (ii) real-world recordings from \textit{UrbanSound8K}~\cite{salamon2014dataset}. Multichannel RIRs are simulated using the image source method~\cite{diaz2021gpurir}. To capture temporal dependencies, we employ $2$-s frame sequences ($K=4$), a duration selected to balance historical context for static sources with rapid detection of directional shifts~\cite{bohlender2021exploiting}.

To model diverse DoA dynamics within each context window, each sample is randomly assigned to one of three motion modes: \textit{static}, \textit{constant-rate}, or \textit{time-varying-rate}. The source moves on the horizontal plane at a fixed radius from the array center, with a randomly chosen initial DoA. Static samples maintain a constant DoA. Constant-rate samples adopt a constant angular velocity sampled from $[-12, 12]^\circ$ per frame. Time-varying-rate samples introduce non-uniform motion through periodic modulation, with amplitude sampled from $[5, 55]^\circ$, random phase, and a cycle count over the context window sampled from $[0.1, 0.2]$. Moving-source signals are generated via time-varying convolution~\cite{yang2024realman}, and the next-frame DoA label is computed analytically from the motion pattern. Since the source speed is assumed to be much less than the speed of sound, the Doppler effect is omitted. 

\begin{table}[!t]
    \centering
    \caption{Configurations of the Datasets.}
    \vspace*{-0.3cm}
    \label{table3}
    \scriptsize
    \begin{tabular}{cc}
        \toprule
        \multicolumn{2}{c}{\textbf{Training and Validation Datasets}} \\
        \midrule
        Noise signal & Synthesized \& real noises \\
        Room size ($\mathrm{m}$) & $\mathrm{R}_1$: $(6,4,3)$, $\mathrm{R}_2$: $(12,8,3.5)$, $\mathrm{R}_3$: $(16,14,4)$ \\
        Array positions & $8$ arbitrary positions in each room \\
        $\mathrm{RT}_{60}$ ($\mathrm{s}$) & $\mathrm{R}_1$: $0.1$, $0.2$, $0.3$; $\mathrm{R}_2$: $0.4, 0.5, 0.6$; $\mathrm{R}_3$: $0.7, 0.8, 0.9$ \\
        SNR (dB) & Uniformly sampled from $10$ to $50$ \\
        \midrule
        \multicolumn{2}{c}{\textbf{Testing Dataset}} \\
        \midrule
        Noise signal & Synthesized \& real noises \\
        Room size ($\mathrm{m}$) & ${\mathrm{R}_1}^\prime$: $(7,5,3)$; ${\mathrm{R}_2}^\prime$: $(11,9,3.2)$; ${\mathrm{R}_3}^\prime$: $(15,13,4.2)$ \\
        Array positions & $4$ arbitrary positions in each room \\
        $\mathrm{RT}_{60}$ ($\mathrm{s}$) & ${\mathrm{R}_1}^\prime$: $0.17$, ${\mathrm{R}_2}^\prime$: $0.48$, ${\mathrm{R}_3}^\prime$: $0.83$ \\
        SNR (dB) & $10$, $20$, $30$, $40$, $50$ \\
        \bottomrule
    \end{tabular}
    \vspace*{-0.2cm}
\end{table}

To enhance the CRNN's robustness, the datasets incorporate diverse RIR variations with different room sizes, array positions, reverberation times ($\mathrm{RT}_{60}$), and signal-to-noise ratio (SNR) levels. For testing, we use noise types and acoustic conditions that are unseen during training. A summary of the dataset configurations is provided in Table~\ref{table3}. In total, the datasets include $86400$ training samples, $9600$ validation samples, and $9600$ test samples per room–SNR subset. 

\begin{table}[!t]
    \centering
    \caption{Classification accuracy of the CRNN.}
    \vspace*{-0.3cm}
    \label{table4}
    \scriptsize
    \begin{tabular}{c ccccc}
        \toprule
        \textbf{Room} & \multicolumn{5}{c}{\textbf{SNR (dB)}} \\
        \cmidrule(lr){2-6}
        & \textbf{10} & \textbf{20} & \textbf{30} & \textbf{40} & \textbf{50} \\
        \midrule
        ${\mathrm R}_1^{\prime}$ & 87.9\% & 90.3\% & 91.3\% & 91.7\% & 91.2\% \\
        ${\mathrm R}_2^{\prime}$ & 86.8\% & 89.9\% & 90.0\% & 90.4\% & 90.2\% \\
        ${\mathrm R}_3^{\prime}$ & 86.9\% & 90.1\% & 90.3\% & 90.3\% & 90.1\% \\
        \bottomrule
    \end{tabular}
    \vspace*{-0.6cm}
\end{table}

\begin{figure*}[!t]
\centering
\includegraphics[width=6.2in]{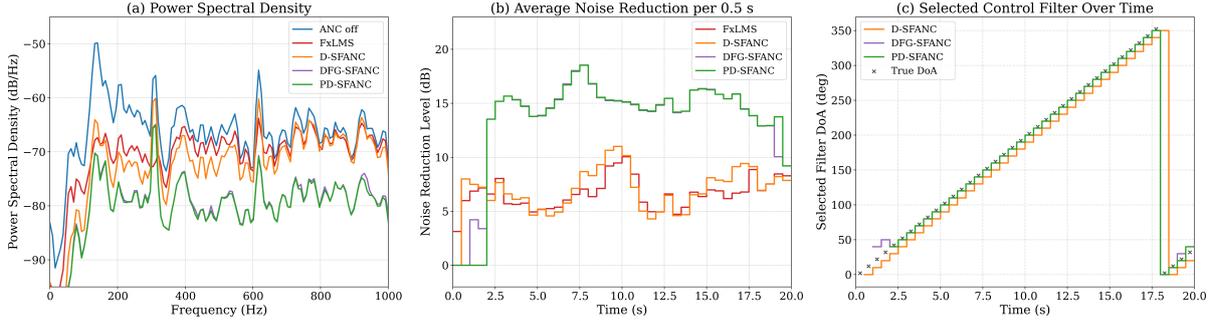}\vspace*{-0.3cm}
\caption{Noise reduction performance in (a) frequency and (b) time domains, and (c) the selected control filter for different ANC methods under vacuum cleaner noise moving at a constant rate, where the DoA varies linearly with an angular velocity of $10^\circ$/$\mathrm{s}$.}
\vspace*{-0.2cm}
\label{fig_4}
\end{figure*}

\begin{figure*}[!t]
\centering
\includegraphics[width=6.2in]{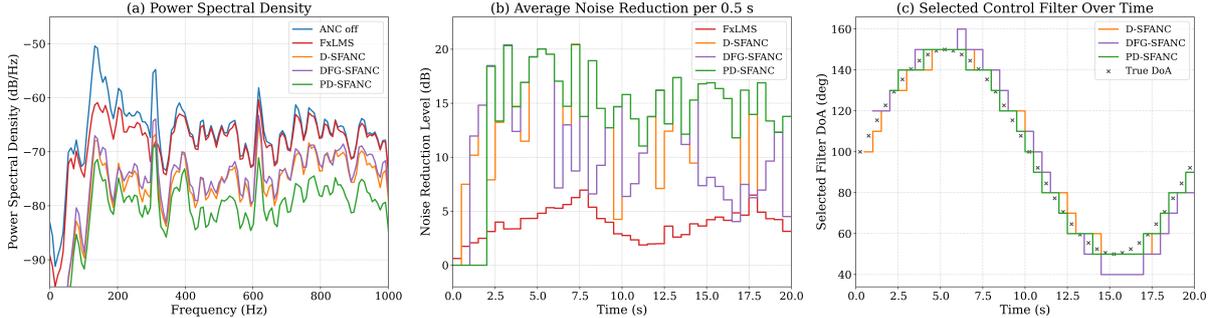}\vspace*{-0.3cm}
\caption{Noise reduction performance in (a) frequency and (b) time domains, and (c) the selected control filter for different ANC methods under vacuum cleaner noise moving at a time-varying rate, where the DoA varies sinusoidally between $50^\circ$ and $150^\circ$.}
\vspace*{-0.5cm}
\label{fig_5}
\end{figure*}

\vspace*{-0.3cm}
\subsection{Effectiveness of the CRNN}
\vspace*{-0.2cm}
The DoA classification accuracy for each room and SNR level is summarized in Table~\ref{table4}, with results averaged across four array positions per room. The proposed CRNN achieves high accuracy across all rooms, exceeding $90$\% at SNRs of $20$ dB and above, with only a slight drop to $87$\% at $10$ dB. These results validate the CRNN’s generalization capability when applied to unseen noise types and acoustic conditions. Moreover, the CRNN is highly efficient, comprising only $0.05$~million parameters and requiring $480.08$~million MACs, which makes it suitable for deployment on resource-constrained co-processors.

\vspace*{-0.3cm}
\subsection{Noise reduction performance}
\vspace*{-0.3cm}
To evaluate the noise reduction performance of PD-SFANC, a rectangular enclosure of size $(11,9,3.2)$ $\mathrm{m}$ is simulated. A multi-reference ANC system is considered with four reference microphones, one secondary source, and one error microphone. The reference microphone array center is located at $(6,4,2.2)$ $\mathrm{m}$, the secondary source at $(4.8,4,2.2)$ $\mathrm{m}$, and the error microphone at $(4.6,4,2.2)$ $\mathrm{m}$. The moving source is simulated via time-varying convolution~\cite{yang2024realman} to move at a fixed radius of $0.4$ $\mathrm{m}$ on the horizontal plane centered at the reference microphone array. Acoustic paths are modeled as RIRs using the image source method~\cite{diaz2021gpurir} with $\mathrm{RT}_{60}$ set to $0.48$ $\mathrm{s}$ and the SNR set to $30$ dB. PD-SFANC is compared against several representative baselines, including FxLMS~\cite{kuo1999active}, D-SFANC~\cite{wangdirectional}, and DFG-SFANC~\cite{su2025co}. For the FxLMS algorithm, the stepsize is set to $1 \times {10^{ - 2}}$ to ensure stability, following the criteria in~\cite{kuo1999active}. D-SFANC and DFG-SFANC employ the same DoA classes as the proposed method. For DFG-SFANC, the adjacent observation node weight is set to $0.02$ and the observation length to $2$, consistent with~\cite{su2025co}. Performance is evaluated using the power spectral density (PSD)~\cite{kuo1999active} and the averaged noise reduction level (NRL) per $0.5$s. The NRL (in dB) is defined as
\begin{equation}
\setlength{\abovedisplayskip}{2pt}
\setlength{\belowdisplayskip}{2pt}
{\mathrm{NRL}} = 10\log_{10} \left[ \sum\nolimits_{n=1}^N d^2(n) / \sum\nolimits_{n=1}^N e^2(n) \right],
\end{equation}
where $d(n)$ is the disturbance at the error microphone, $e(n)$ is the residual error, and $N=8000$ denotes the evaluation window length, which is aligned with the control filter update rate.

Two experiments are conducted using real-world vacuum cleaner noise under constant-rate and time-varying-rate source movement scenarios, respectively. In the first experiment, the source DoA is designed to linearly increase from $0^\circ$ with a constant angular velocity of $10^\circ/\mathrm{s}$ for $20$ $\mathrm{s}$. Fig.~\ref{fig_4}(a) and (b) illustrate the PSD of the error signals and the NRL over time for the four comparative algorithms. The control filters selected by D-SFANC, DFG-SFANC, and PD-SFANC during this motion are depicted in Fig.~\ref{fig_4}(c). The results indicate that while all methods can adapt to the moving source, D-SFANC exhibits a one-frame lag due to its lack of predictive capability. In contrast, both DFG-SFANC and PD-SFANC select filters that align more closely with the true source DoA by exploiting temporal context. Consequently, both DFG-SFANC and PD-SFANC maintain an NRL above $15$ dB for most of the duration, while D-SFANC produces a lower NRL with high-amplitude fluctuations caused by delayed filter switching. Meanwhile, the FxLMS algorithm exhibits limited noise reduction performance, as it requires a longer convergence time, whereas the SFANC-based methods can update the filter at the frame level.

In the second experiment, the source DoA follows a sinusoidal trajectory between $50^\circ$ and $150^\circ$ for $20$ $\mathrm{s}$. As shown in Fig.~\ref{fig_5}(a) and (b), PD-SFANC demonstrates superior performance, maintaining stable and high noise reduction throughout the trajectory due to the effective filter pre-selection enabled by DoA prediction. In contrast, FxLMS and D-SFANC achieve lower NRLs with greater fluctuations, as they fail to adapt sufficiently fast to the directional changes. Notably, DFG-SFANC exhibits significant performance drops at specific intervals, such as around the $7$-th and $15$-th $\mathrm{s}$. This suggests that DFG-SFANC struggles to track sources with rapidly varying acceleration, particularly in reverberant environments, whereas PD-SFANC demonstrates more robust tracking capabilities.

\vspace*{-0.45cm}
\section{Conclusion}
\vspace*{-0.25cm}
\label{section4}
This paper presents a novel PD-SFANC method for handling moving noise sources. By leveraging temporal directional dependencies across multiple consecutive frames, the CRNN predicts the source DoA and proactively selects the most suitable control filter. Numerical simulations demonstrate that the proposed CRNN achieves accurate DoA predictions and exhibits robust generalization to unseen noise types and acoustic environments. Furthermore, comparative evaluations against representative ANC baselines confirm the superiority of PD-SFANC, demonstrating robust noise reduction and rapid tracking response across different source movement scenarios.


\newpage
\bibliographystyle{IEEEtran}
\bibliography{mybib}

\end{document}